\documentclass[twocolumn,english,aps,pra,10pt,superscriptaddress,floatfix]{revtex4-2}

\usepackage{times}
\usepackage{bm}
\usepackage{graphicx}
\usepackage{amssymb,amsfonts,amsmath,amsbsy,bm,t1enc,latexsym}
\usepackage{float}
\usepackage{siunitx}
\usepackage{xcolor}
\definecolor{fancyblue}{RGB}{70, 120, 200}     
\definecolor{fancymagenta}{RGB}{200, 50, 130}  
\usepackage[colorlinks=true,
citecolor=fancyblue,
linkcolor=fancymagenta]{hyperref}

\usepackage[english]{babel}
\usepackage{url}%

\usepackage{color,soul}
\usepackage{textcomp}
\usepackage[]{lipsum}

\usepackage[markup=underlined, defaultcolor=blue]{changes} 
\usepackage{todonotes}

\usepackage{titlesec}
\titlespacing*{\subsection}
{0pt}{0.8em}{0.3em} 

\begin{document}
\title{Watt-level second harmonic generation in periodically poled thin-film lithium tantalate} 

\author{Nikolai Kuznetsov}
\thanks{These authors contributed equally to this work.}
\affiliation{Institute of Physics, Swiss Federal Institute of Technology Lausanne (EPFL), CH-1015 Lausanne, Switzerland}

\author{Zihan Li}
\thanks{These authors contributed equally to this work.}
\affiliation{Institute of Physics, Swiss Federal Institute of Technology Lausanne (EPFL), CH-1015 Lausanne, Switzerland}

\author{Tobias J. Kippenberg}
\email[]{tobias.kippenberg@epfl.ch}
\affiliation{Institute of Physics, Swiss Federal Institute of Technology Lausanne (EPFL), CH-1015 Lausanne, Switzerland}
\affiliation{Center of Quantum Science and Engineering (EPFL), CH-1015 Lausanne, Switzerland}
\affiliation{Institute of Electrical and Micro Engineering (IEM), Swiss Federal Institute of Technology Lausanne (EPFL), CH-1015 Lausanne, Switzerland}
	
\maketitle

\textbf{
Second-harmonic generation (SHG)~\cite{franken_generation_1961} is a fundamental tool in modern laser technology, enabling efficient coherent frequency conversion to remote optical bands, serving as the basis for self-referencing of femtosecond lasers, and being essential for the generation of quadrature-squeezed light, as used in LIGO.
State-of-the-art frequency conversion relies on bulk crystals and ridge waveguides, although continuous-wave (CW) conversion efficiency in bulk crystals is limited by short interaction lengths and large mode areas.
Ridge waveguides offer better performance with lower pump power requirements, yet must span several centimeters to deliver high output power~\cite{pecheur_watt-level_2021, carpenter_cw_2020, umeki_highly_2010}, complicating fabrication and narrowing the bandwidth.
Recently, the ability to achieve SHG in thin-film lithium niobate integrated photonic circuits has attracted significant interest, offering several-orders-of-magnitude improvement in SHG under CW pumping due to the stronger optical mode confinement~\cite{zhu_integrated_2021, li_high_2023, wang_lithium_2024}.
Quasi-phase-matched~\cite{fejer_quasi-phase-matched_1992} periodically poled thin-film lithium niobate waveguides have shown enhanced efficiency at low pump powers~\cite{wang_ultrahigh-efficiency_2018}.
However, lithium niobate~\cite{kong2020recent} has a low optical damage threshold, even in MgO-doped substrates, which limits SH power output to well below the watt level.
Here, we overcome this challenge and demonstrate \SI{7}{\milli\meter}-long periodically poled thin-film lithium tantalate (PPLT) waveguides that achieve high SH output in the CW regime, with generated power exceeding \SI{1}{\watt} and off-chip output above \SI{0.5}{\watt} at \SI{775}{\nano\meter} under \SI{4.5}{\watt} pump power -- limited only by our experimental setup.
Lithium tantalate offers a higher optical damage threshold~\cite{yan_high_2020, sinha_room-temperature_2008, suntsov_optical_2024} than lithium niobate and supports watt-level operation.
By optimizing electrode geometry and poling conditions, we achieved reproducible poling despite lithium tantalate's coercive field being nearly four times higher than that of MgO-doped lithium niobate.
Although its effective nonlinearity is more than five times lower, we achieve watt-level CW output with a short waveguide.
This preserves the bandwidth and demonstrates the potential of PPLT photonic circuits for high-power applications in integrated lasers~\cite{siddharth_piezoelectrically_2024}, quantum photonics, AMO physics, optical clocks, and frequency metrology.}

Second harmonic generation, first discovered in 1961~\cite{franken_generation_1961}, is a widely used nonlinear process in science and technology.
It is the basis for applications ranging from simple laser pointers to f-2f self referencing for optical frequency combs and the generation of low-noise visible and near-UV laser emission from more mature optical bands.
In recent decades, a variety of second-order nonlinear materials have been developed, including lithium niobate (LiNbO$_3$), lithium tantalate (LiTaO$_3$), and other compounds~\cite{shoji_absolute_1997} such as potassium niobate (KNbO$_3$), potassium dihydrogen phosphate (KH$_2$PO$_4$, KDP), potassium titanyl phosphate (KTiOPO$_4$, KTP), beta barium borate (BaB$_2$O$_4$, BBO), and lithium triborate (LiB$_3$O$_5$, LBO).
Essential to efficient frequency doubling is the method of quasi-phase-matching (QPM)~\cite{fejer_quasi-phase-matched_1992}, which involves periodic inversion of the material’s domains to counteract phase mismatch and enable continuous SH power build-up. 
QPM is typically implemented through periodic poling, in which ferroelectric domains are permanently inverted using an external electric field applied via comb-shaped electrodes on the crystal surface.  

\begin{figure*}[htb!]
	\centering
	\includegraphics[width=1\textwidth]{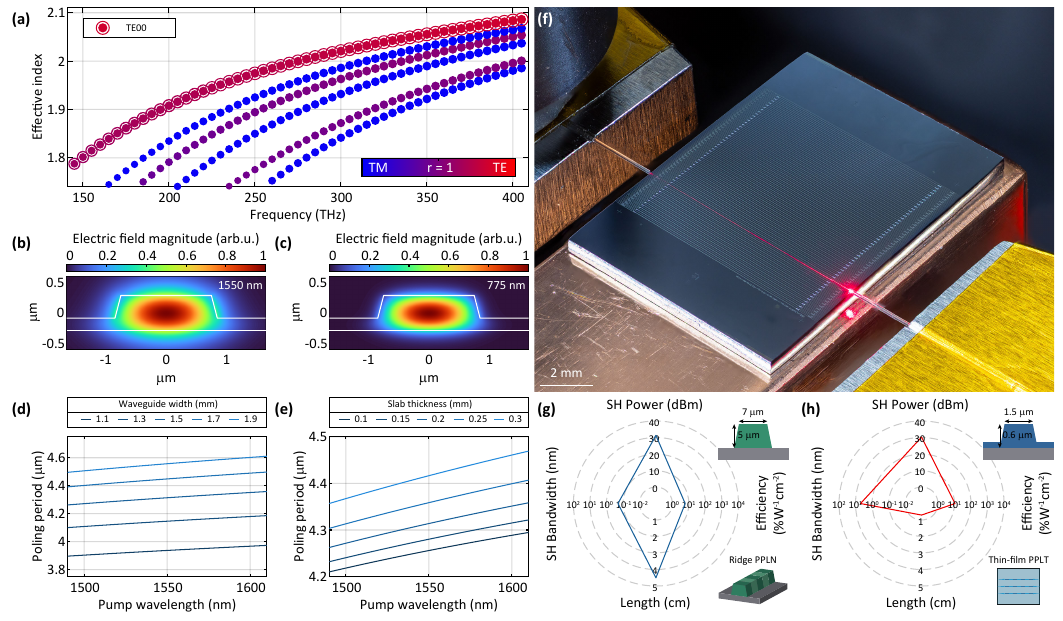}
	\caption{\textbf{Second harmonic generation in periodically poled thin-film lithium tantalate waveguides.} 
		\textbf{(a)}~Numerical simulations of the effective index of the first six modes in the LiTaO$_3$ waveguide reported in this work.
		The data for the fundamental TE mode is highlighted.
		Colors indicate the polarization ratio of each mode.
		\textbf{(b,~c)}~Mode field profiles for pump and SH waves, respectively.
		\textbf{(d,~e)}~Dependence of the poling period on the waveguide width (the slab thickness is fixed) and on the slab thickness (the waveguide width is fixed), respectively; the waveguide height is fixed in both cases.
		\textbf{(f)}~A focus-stacked image of a PPLT chip emitting red light produced by the SHG process in a thin-film PPLT waveguide.
		A PPLT chip, designed for a pump wavelength of \SI{1307}{\nano\meter}, is presented in this image, as emission at \SI{775}{\nano\meter} is already cut by filters preinstalled in modern cameras and cannot be captured.
		\textbf{(g, h)}~Performance comparison of a ridge PPLN waveguide (based on the data from Ref.~\cite{kashiwazaki_high-gain_2021}) and the thin-film PPLT waveguide from this work, respectively.
	}
	\label{fig:intro}
\end{figure*}

It has been shown that this technique can be used for photonic integrated circuits (PICs) based on thin-film ferroelectric materials~\cite{zhu_integrated_2021, li_high_2023, wang_lithium_2024, wang_ultrahigh-efficiency_2018, lu_periodically_2019} (Fig.~\ref{fig:intro}(f)).
These PICs enhance conversion efficiency by several orders of magnitude compared to bulk crystals and ridge waveguides due to significantly smaller optical mode areas.
Recent studies on periodically poled LiNbO$_3$ (PPLN) have demonstrated impressive performance~\cite{chen_adapted_2024, xin_wavelength-accurate_2025, franken_milliwatt-level_2025}, including more than \SI{80}{\%} absolute conversion at around \SI{20}{\milli\watt} of pump power.
Other methods relying on modal phase-matching have recently been shown to achieve efficient second-order nonlinear interaction~\cite{zheng_nonlinear_2021, luo_seminonlinear_2019, luo_highly_2018, shi_efficient_2024, hefti_fabrication-tolerant_2025}.
Moreover, a substantial amount of work over the past decade has led to important advances in poled ferroelectric PIC-based devices, enabling $\chi^2$-enhanced supercontinuum generation~\cite{jankowski_ultrabroadband_2019, cheng_continuous_2024, ludwig_ultraviolet_2024}, self-referencing of optical frequency combs~\cite{okawachi_chip-based_2020},  microcomb generation~\cite{bruch_pockels_2021, tang_broadband_2024, stokowski_integrated_2024}, and optical quadrature squeezing~\cite{park_single-mode_2024, stokowski_integrated_2023}, extending the functionality of integrated photonics.
While LiNbO$_3$ is an attractive material for photonic applications, its strong photorefractive response and low optical damage threshold~\cite{kong2020recent} limit its use in high-power regimes despite the high conversion efficiency, and the highest reported values of the SH power are usually within tens of milliwatts~\cite{chen_adapted_2024}, rarely exceeding \SI{100}{\milli\watt}~\cite{wang_ultrahigh-efficiency_2018} -- below what can be achieved in ridge waveguides. 

Recently, lithium tantalate (LiTaO$_3$) photonic integrated circuits have been demonstrated~\cite{wang_lithium_2024}.
Compared with LiNbO$_3$, this material exhibits similar electro-optical coefficient, wider transparency window, and low bias drift in modulators~\cite{wang_ultrabroadband_2024}.
Important to our work here is that it possesses, contrary to LiNbO$_3$, a higher optical damage threshold even without MgO doping~\cite{sinha_room-temperature_2008, suntsov_optical_2024}.
As LiTaO$_3$ is a low-birefringent material, mode-crossings are rare and typically not present for the fundamental modes in typical waveguide cross-sections (Fig.~\ref{fig:intro}(a)).
Therefore, the same waveguide cross-section can be used to produce SH emission at different wavelengths, simplifying the circuit design (Fig.~\ref{fig:intro}(f)).
However, periodically poled lithium tantalate (PPLT) PICs are currently only in their early development stages~\cite{chen_periodic_2023, chen_continuous-wave_2025}, and so far have not demonstrated superior performance to LiNbO$_3$. 

Here we overcome this challenge and achieve, for the first time, watt-level output.
In this work, we develop high-quality periodically poled thin-film LiTaO$_3$ waveguides and demonstrate continuous-wave (CW) SH output power exceeding \SI{1}{\watt}, with directly measured off-chip SH power above \SI{0.5}{\watt} at \SI{775}{\nano\meter} wavelength.
In contrast, MgO-doped LiNbO$_3$ wafers help to mitigate the photorefractive effect, but to the best of our knowledge, thin-film integrated circuits based on this material cannot sustain prolonged watt-level operation at near-visible or visible wavelengths.

We show that PPLT waveguides can withstand \SI{4.5}{\watt} of off-chip pump power at \SI{1550}{\nano\meter}, and achieve an SH power output comparable to the power produced in state-of-the-art PPLN ridge waveguides (Fig.~\ref{fig:intro}(g)) which are capable of handling higher powers due to their larger waveguide cross-section.
However, despite the higher material nonlinearity, generally ranging from \SI{20}{\pico\meter\volt^{-1}} to \SI{30}{\pico\meter\volt^{-1}} at different wavelengths, as reported in the literature for bulk compounds~\cite{shoji_absolute_1997}, PPLN ridge waveguides can only achieve high SH power by using several-centimeter-long waveguides, which is also the reason for SH bandwidth narrowing.
Given that periodically poled sections can be integrated on the same photonic circuit with other components -- such as low-birefringent microresonators~\cite{wang_lithium_2024}, arrayed waveguide gratings~\cite{hulyal_arrayed_2025}, and electro-optic modulators~\cite{wang_integrated_2018, zhang_broadband_2019, zhang_ultrabroadband_2025, wang_ultrabroadband_2024} -- our results extend the capabilities of the recently emerged LiTaO$_3$ integrated photonic platform towards on-chip optical parametric oscillators, quadrature-squeezed light sources, and visible lasers.

\subsection*{Periodic poling of LTOI photonic integrated waveguides}
\begin{figure*}[htb!]
	\centering
	\includegraphics[width=1\textwidth]{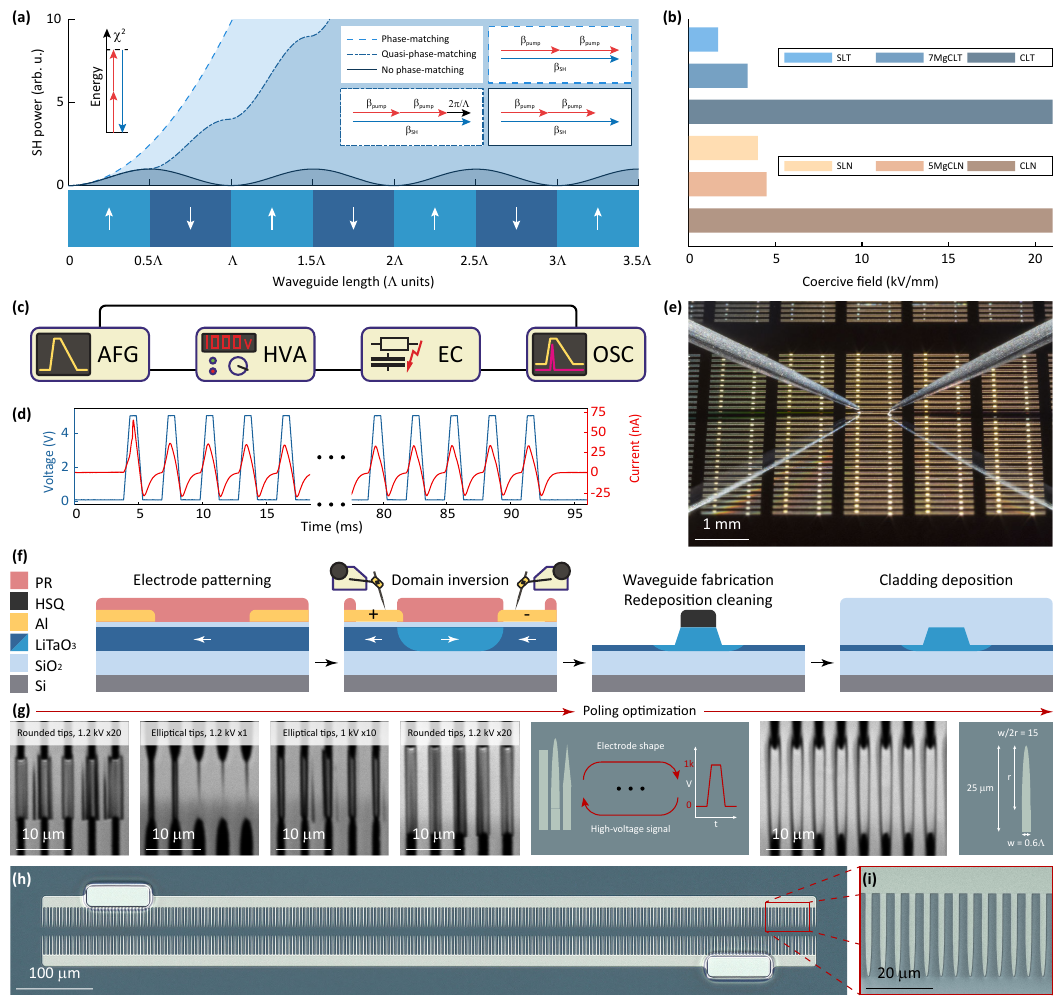}
	\caption{\textbf{Periodic poling of thin-film LiTaO$_3$.} 
		\textbf{(a)}~Type 0 SHG process in crystals or waveguides made of non-centrosymmetric materials with different phase-matching conditions.
		Inset on the left -- energy conservation schematic in the SHG process.
		Insets on the right -- momentum conservation conditions for different types of phase-matching.
		$\Lambda$ is a single QPM period.
		\textbf{(b)}~Comparison of typical coercive field magnitudes reported in literature for bulk stoichiometric (S), congruent (C), and MgO-doped congruent (xMgC) compounds of LiNbO$_3$ (LN) and LiTaO$_3$ (LT) at room temperature~\cite{kim_coercive_2002, ishizuki_periodical_2003, ishizuki_mg-doped_2008}.
		\textbf{(c)}~A schematic of the setup for periodic poling. 
		AFG: arbitrary function generator.
		HVA: high-voltage amplifier.
		EC: electric circuit.
		OSC: oscilloscope.
		\textbf{(d)}~A typical signal recorded on the oscilloscope during poling, indicating a sequence of applied poling pulses and the collected current produced by domain inversion and the strain circuit capacitance.
		\textbf{(e)}~A focus-stacked image of a set of test poling electrodes, and two electric probe needles brought into a contact with the electrode pads.
		\textbf{(f)}~Four main steps of the waveguide fabrication process, and the material stack of the sample at each step.
		First, the electrodes are patterned and entire chip is covered with photoresist; in the next step, the pads for probes are opened and poling is performed.
		Finally, the waveguides are fabricated using dry etching and, optionally, cladded with silicon dioxide.
		\textbf{(g)}~Two-photon microscope images showing a few domain-inverted structures obtained, among numerous others, during poling optimization which includes the optimization of both the electrode shape and the high-voltage poling signal.
		First images show chaotic inversion, and the rightmost image demonstrates one of the best poling results that can be routinely achieved after our poling technique has been optimized. 
		\textbf{(h)}~An optical microscope image of comb-shaped electrodes for periodic poling.
		A testing \SI{1}{\milli\meter}-long electrode is shown; \SI{7}{\milli\meter}-long electrodes were used for the actual waveguides -- no dependence on the electrode length is observed for periodic poling with the optimized conditions.
		The edges of the photoresist layer are visible along the boundaries of the pads where the photoresist is removed.
		\textbf{(i)}~A false-colored SEM image of a part of the same electrode shown in \textbf{(h)}. 
		}
	\label{fig:poling}
\end{figure*}

To enable efficient SHG in thin-film ferroelectric circuits, the waveguide must have a uniform structure with periodically inverted nonlinearity to support unidirectional power transfer from the pump to the SH wave. (Fig.~\ref{fig:poling}(a)).
We find that, despite the high coercive field of \SI{21}{\kilo\volt/\milli\meter} reported for bulk LiTaO$_3$ (Fig.~\ref{fig:poling}(b))~\cite{kim_coercive_2002, ishizuki_periodical_2003, ishizuki_mg-doped_2008}, domain inversion in thin-film substrates is achievable using poling techniques similar, although not identical, to those developed for thin-film PPLN.

The poling process proved sensitive to both the electrode shape and the poling signal (Fig.~\ref{fig:poling}(g)), and we needed to fabricate several thousand electrodes for the optimization of our poling routines.
As depicted in Fig.~\ref{fig:poling}(f), we start the fabrication of PPLT circuits using an X-cut LiTaO$_3$ chip, which consists of \SI{576}{\nano\meter}-thick LT film and \SI{4.7}{\micro\meter} SiO$_2$ insulation layer.
First, we deposit a \SI{100}{\nano\meter}-thick layer of SiO$_2$ using plasma-enhanced chemical vapor deposition (Oxford PlasmaLab 100) on the LT surface to prevent possible leakage current.
Next, we fabricate \SI{100}{\nano\meter} thick comb-shaped aluminum electrodes for periodic poling (Fig.~\ref{fig:poling}(h,i)) by the lift-off method using a bi-layer resist of PMMA/MMA and E-beam evaporation (Alliance Concept EVA 760).
Since the air breakdown voltage is approximately \SI{3}{\kilo\volt/\milli\meter}, we protect the comb-shaped electrodes by covering the entire sample with \SI{2}{\micro\meter}-thick photoresist (AZ ECI 3027), opening the probe pads afterwards.
These pads are placed at least \SI{800}{\micro\meter} away from each other.
This approach has proven reliable and more user-friendly than using silicon oil, which has been employed in some previous studies~\cite{nagy_reducing_2019}.
Before proceeding to the waveguide fabrication steps, we apply a short high-voltage (HV) pulse to the electrodes using HV DC probes (Fig.~\ref{fig:poling}(e)).
We work with separate chips to increase flexibility in process optimization, speed up fabrication, and reduce material waste.
In principle, these methods can be applied at the wafer scale using deep-ultraviolet lithography for electrode and waveguide patterning.
We program an arbitrary function generator (AFG, Tektronix AFG31102) to produce the desired pulse shape with the amplitude below \SI{10}{\volt} (Fig.~\ref{fig:poling}(c)).
The output is split: one part is sent to an oscilloscope (Rohde\&Schwarz RTA4004), and the other is fed into a high-voltage amplifier (HVA, Trek 2200; with a fixed gain of \SI{200}{\volt/\volt}).
We note that the large-signal bandwidth of our HVA is below 10~kHz, which currently prevents us from exploring the poling techniques with shorter pulses.
The HVA output is delivered to the HV probes via a custom electric circuit (assembled similarly to the circuit in Ref.~\cite{nagy_reducing_2019}), which collects the signal between the probes and sends it to the oscilloscope while also providing protection in case of breakdown.
Before testing the PPLT waveguides optically, we optimized our poling routine on test samples fabricated under identical conditions.
We find that both the electrode tip shape and the pulse waveform significantly affect poling performance.
We have studied various shapes of the electrodes, from rectangular to triangular and rounded tips.
After multiple iterations, we have converged to elliptical electrode shape, as it produced the best results (Fig.~\ref{fig:poling}(g)).
Unlike in PPLN, we find that single-pulse poling is not reliable for domain inversion in PPLT.
Instead, we use a sequence of 20–30 pulses, similar to poling techniques shown in Refs.~\cite{chen_periodic_2023, chen_continuous-wave_2025}.
Our current configuration is not unique -- the parameter space is broad, and alternative electrode designs and pulse shapes may also yield high-quality poling.
In this work, we use electrodes with elliptically shaped \SI{25}{\micro\meter}-long tips (axis ratio 1:15), a duty cycle of \SI{40}{\%}, and a \SI{10}{\micro\meter} gap between the electrodes.
A \SI{15}{\micro\meter} gap also shows a good poling, as shown in the rightmost panel in Fig.~\ref{fig:poling}(g), at the cost of slightly increased voltage, \SI{1.15}{\kilo\volt}.
As expected, the electrode duty cycle influences the width of inverted domains, though its importance is reduced by the ability to fine-tune domain width via voltage.
We observe no significant variation for duty cycles between \SI{40}{\%} and \SI{60}{\%}.
Each pulse includes a \SI{0.5}{\milli\second} rise, \SI{0.5}{\milli\second} flat, and \SI{0.5}{\milli\second} fall segment.
The peak voltage is set to \SI{1}{\kilo\volt}, and the interval between pulses is \SI{1.5}{\milli\second}.
A typical pulse sequence is shown in Fig.~\ref{fig:poling}(d).
The blue line indicates the signal sent to the HVA, and the red line shows the poling current between probes -- domain inversion induces charge redistribution within the material, producing this signal.
The first pulse initiates the inversion, generating most of the current, while subsequent pulses primarily assist the lateral growth of the inverted domains, each contributing only marginally.
However, this current is largely masked by stray capacitance in the circuit, which results in the oscillating signal following the poling pulse sequence.
We also observe that a minimum distance must be maintained between pairs of poling electrodes on the same chip to avoid “poling cross-talk” -- the poling quality between one pair of electrodes can be influenced by the conditions (e.g., voltage, pulse duration) used for an adjacent pair (more information on this is presented in the Supplementary Material).
This effect was never observed in our PPLN poling experiments (to be reported elsewhere) and may be attributed to charge accumulation on neighboring electrodes.
Further studies are required to fully understand this phenomenon, and more details on the periodic poling process optimization are presented in the Supplementary Materials for this work.
We find that placing electrodes at least \SI{50}{\micro\meter} apart eliminates this issue.
After poling, we remove the photoresist and inspect the domain inversion using a two-photon microscope (Zeiss LSM 710 NLO).
Under identical imaging conditions, PPLT samples appear dimmer than PPLN samples of the same thickness -- possibly indicating lower intrinsic nonlinearity or crystal quality.

We then proceed to circuit fabrication following routines similar to those described in Ref.~\cite{wang_lithium_2024}.
Before the waveguide fabrication, we strip the protective photoresist, Al electrode, and SiO$_2$ insulation layer by step-by-step wet etching, with acetone, TechniEtch Al80, and buffered oxide etch solution, respectively.
Then, a hydrogen silsesquioxane (HSQ) layer of around \SI{900}{\nano\meter} is coated on the surface as the etching mask.
We expose the waveguides pattern by \SI{100}{\kilo\volt} electron-beam-lithography (Raith EBPG5000).
After development with \SI{25}{\%} tetramethylammonium hydroxide (TMAH) solution, we transfer the HSQ pattern to the LT layer by argon ion beam etching (Veeco Nexus IBE350).
A precise etching depth can be achieved by multi-step etching and calibration.
To reduce the roughness on the waveguide sidewall, which is caused by the redeposition during the dry etching, a following wet etching step based on the high-temperature (\SI{80}{\celsius}) potassium hydroxide (KOH) solution is applied.
The second E-beam lithography and dry etching step creates a double-layer taper to improve the facet fiber-to-chip coupling efficiency~\cite{he_low-loss_2019}.
Next, we clad the PPLT waveguides by an \SI{1.5}{\micro\meter} hydrogen-free SiO$_2$ layer by inductively-coupled-plasma chemical vapor deposition (ICP-CVD, Oxford PlasmaPro 100)~\cite{qiu_hydrogen-free_2024}.
Finally, we singulate the chip by consecutively dry etching SiO$_2$, deep Si etching, and cleavage.

\subsection*{Waveguide characterization}

\begin{figure*}[htb!]
	\centering
	\includegraphics[width=1\textwidth]{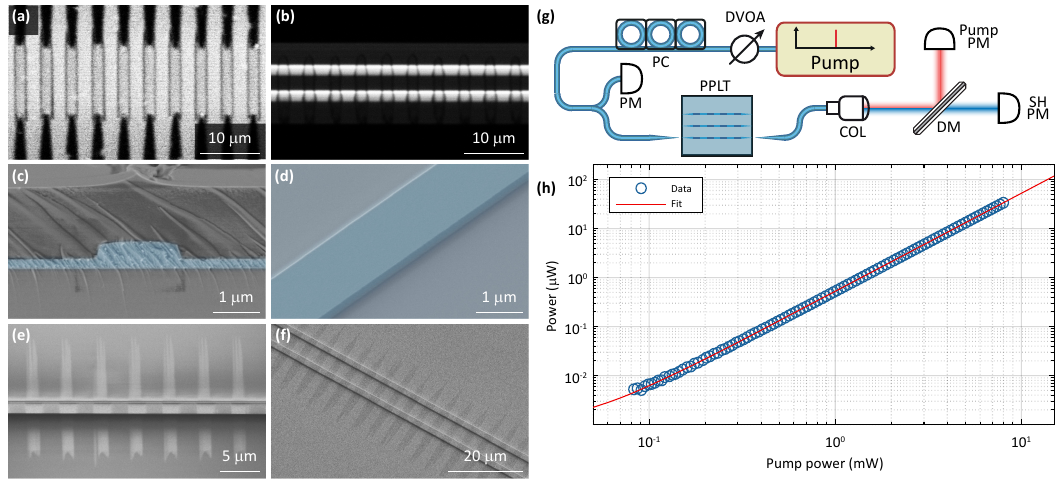}
	\caption{\textbf{Inspection of thin-film PPLT waveguides and SHG efficiency measurements.} 
		\textbf{(a)}~A two-photon microscope image of inverted domains before the waveguide fabrication.
		\textbf{(b)}~Same as \textbf{(a)}, but after the waveguide fabrication was complete.
		\textbf{(c)}~A false-colored SEM image of the waveguide cross-section.
		\textbf{(d)}~A false-colored SEM image of the waveguide sidewalls, demonstrating the absence of etching-induced corrugation or roughness.
		\textbf{(e)}~An SEM image of a waveguide that allows fast high-resolution observation of periodic domain inversion.
		\textbf{(f)}~Same as \textbf{(e)}, but with a different sample, a different angle, and different magnification.
		\textbf{(g)}~Experimental setup used to measure the SHG efficiency.
		DVOA: digital variable optical attenuator.
		PC: polarization controller.
		COL: collimator.
		PM: powermeter.
		\textbf{(h)}~SH power measured and calibrated after varying the pump power.
		Efficiency is extracted from the slope of the red solid curve.
	}
	\label{fig:shg}
\end{figure*}

We fabricate multiple waveguides on the same chip, with widths of \SI{1.5}{\micro\meter} and \SI{1.6}{\micro\meter}, and implement several poling periods, varied in \SI{5}{\nano\meter} steps around a target value of $\Lambda$~=~\SI{4308}{\nano\meter} (Fig.~\ref{fig:shg}(a)).
The average sample thickness is \SI{576}{\nano\meter}, and etching is stopped early to leave a \SI{200}{\nano\meter}-thick slab.

We inspect the sample quality at various fabrication stages.
As described above, domain inversion is verified using a two-photon microscope, which, in this case, functions as an SH microscope.

SH photons generated near the domain walls interfere destructively, creating dark lines in the image
We optimize the poling routine to produce a uniform, high-contrast pattern with an approximate \SI{50}{\%} duty cycle (Fig.~\ref{fig:shg}(a)).
Post-waveguide fabrication, we perform the same imaging to confirm that no domain back-switching has occurred (Fig.~\ref{fig:shg}(b)).
Using a scanning electron microscope (SEM), we confirm that waveguide dimensions match the design parameters (Fig.~\ref{fig:shg}(c)) and exhibit smooth sidewalls, free of corrugation from anisotropic etching (Fig.~\ref{fig:shg}(d)).
We also find that conventional SEM imaging -- without modification -- provides unexpectedly high-resolution inspection of domain inversion.
By adjusting the angle between the sample and the detector, inverted domains become clearly visible both on the waveguides and the slab (Fig.~\ref{fig:shg}(e,~f)).
We emphasize that this pattern is not due to surface topography, as it disappears from other viewing angles.
Because the contrast is visible on the thin slab, this imaging technique allows for non-destructive estimation of the minimum poling depth.
Compared to piezoresponse force microscopy, which also resolves domain structures, SEM imaging is significantly faster, more convenient, and suitable at any fabrication step -- before or after waveguide patterning.
Since periodic poling induces no major structural changes apart from permanent ion displacement within the crystal lattice, we attribute the observed SEM contrast to slight variations in surface electron scattering conditions between inverted and non-inverted domains.

After completing microscopic characterization, we proceed to optical measurements of low-power SHG efficiency.
SHG performance depends on both material properties and waveguide design.
Waveguide geometry is chosen to ensure strong confinement and mode overlap while maintaining a sufficiently large poling period to facilitate fabrication.
To measure SHG efficiency, we use the experimental setup shown in Fig.~\ref{fig:shg}(g).
We use a continuously-tunable laser (TOPTICA CTL) and a digital variable optical attenuator (DVOA, OZ OPTICS DA-100) to acquire multiple data points.
We optimize the pump wavelength, fiber coupling, and polarization to maximize the SHG power.
Then the DVOA is programmed to change the pump attenuation from \SI{30}{\dB} to \SI{0}{\dB} with a step of \SI{0.2}{\dB}.
At each step, input pump power and output SH power values are measured and recorded (powermeters Thorlabs S144C and S120C, respectively).
Pump coupling and SH collection are achieved using inverse tapers and lensed fibers.
While pump coupling efficiency can be accurately calibrated (assuming identical input and output facet losses of approximately \SI{4}{\dB} per facet), incorrect estimation of the SH coupling loss at the output facet can lead to an overestimated efficiency.
To address this, we perform high-power measurements (described below), calibrate input and output pump power, and fit the SH coupling coefficient to satisfy energy conservation.
After extracting the SH coupling coefficient from high-power data, we use it to calibrate low-power efficiency measurements.
To close the loop, we input the fitted efficiency into numerical simulations of the SHG process, which show excellent agreement with experimental results.
This method yields more reliable efficiency values than direct loss measurements and avoids overestimation.
Our measured efficiency is \SI{107}{\%\watt^{-1}\cm^{-2}} (Fig.~\ref{fig:shg}(h)) -- evidently lower than typical values reported for LiNbO$_3$, and somewhat below theoretical estimations based on the values of nonlinear coefficient of bulk LiTaO$_3$.
Although both materials exhibit similar electro-optic responses~\cite{wang_lithium_2024, wang_ultrabroadband_2024}, the values of second-order optical susceptibility can be notably different as measured in bulk crystals and reported in earlier literature~\cite{shoji_absolute_1997}.
The measurements in Ref.~\cite{shoji_absolute_1997} already suggest the reduction of nonlinearity by almost a factor of~3; this can, however, be only part of the reason behind low efficiency because our data shows additional reduction by a factor of~2.
Although domain inversion in our PPLT waveguides appears satisfactory under inspection, the precise reason for the reduced efficiency remains unclear.
Improving the crystal growth routines and optimizing the waveguide fabrication methods can potentially enhance the efficiency.
However, we currently cannot directly assess the crystal quality or effective material nonlinearity by other techniques -- both of which may differ in thin films compared to bulk crystals.
Some degradation of efficiency may be due to thin-film thickness non-uniformity.
Our measurements show certain widening of the SHG spectra, as shown below, which is an indirect indication of a sample nonuniformity, and our PPLT waveguides are already long enough to be affected by thickness variations. 
Precise thickness map measurements before fabrication could be helpful, and further investigation is required to resolve these questions.

\subsection*{Watt-level continuous-wave second harmonic}

\begin{figure*}[htb!]
	\centering
	\includegraphics[width=1\textwidth]{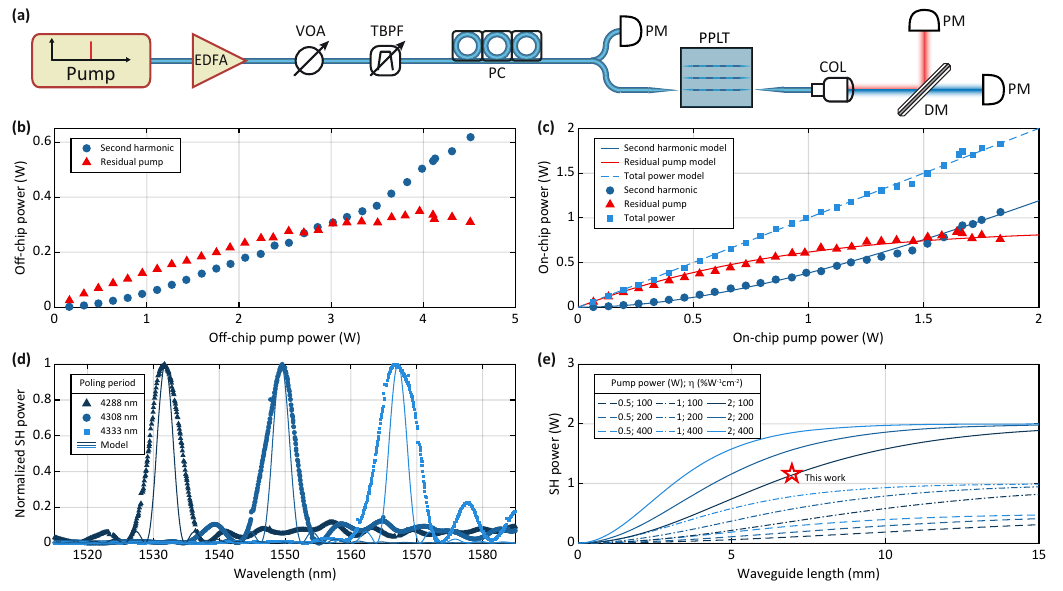}
	\caption{\textbf{Generation and measurement of watt-level CW SH emission.} 
		\textbf{(a)}~Extended experimental setup for high-power CW optical pumping.
		EDFA: erbium-doped fiber amplifier.
		TBPF: tunable bandpass filter.
		VOA: variable optical attenuator.
		PC: polarization controller.
		COL: collimator.
		DM: dichroic mirror.
		PM: powermeter.
		\textbf{(b)}~Direct off-chip power measurements of SH and residual pump.
		\textbf{(c)}~Fully-calibrated on-chip power measurements of SH and residual pump, and results of numerical modeling (solid lines).
		\textbf{(d)}~SH spectra measured in different waveguides with varied poling periods.
		\textbf{(e)}~Numerical simulations of potential SHG performance with improved efficiency in waveguides of different lengths under varying pump powers.
		The red star indicates the results in this work.
	}
	\label{fig:highpower}
\end{figure*}

To perform high-power measurements, we modify our setup by installing an erbium-doped fiber amplifier (EDFA, Keopsys CEFA-C), an optical attenuator (Sch\"after+Kirchhoff 48AT-0), and a bandpass filter (Agiltron FOTF), as shown in Fig.~\ref{fig:highpower}(a).

A \SI{1}{\%} portion of the pump power is tapped to a calibrated powermeter (Thorlabs S144C).
When increasing the pump power, at each pump power level, we optimize input and output fiber coupling, as well as pump polarization.
The pump wavelength is tuned to maximize SH output.
Following the SH power measurement (powermeter -- Thorlabs S140C), we re-optimize the output fiber position to collect the maximum residual pump (powermeter -- Thorlabs S148C), ensuring that the input polarization and wavelength remain consistent.
Re-optimization is necessary as the optimal position depends on the wavelength -- pump and SH waves have different spot sizes, and beams diverge differently after the waveguide.
We use a collimator and free-space optical components to measure the output power.
Two beams at different wavelengths cannot be collimated simultaneously, and to avoid beam divergence after the collimator, we make free-space optical paths as short as possible and ensure that all emission is collected by the powermeters.

We directly measure more than \SI{0.5}{\watt} of SH emission after the dichroic mirror (Fig.~\ref{fig:highpower}(b)).
We do not further increase the pump power, as our SH powermeter is limited to \SI{0.5}{\watt}, and we choose not to modify the setup because the pump power is already approaching the output limit of our EDFA.
After calibrating the dichroic mirror losses, we estimate the directly collected SH power to be approximately \SI{0.6}{\watt}.
Using the full system calibration described earlier, we estimate the on-chip generated SH power to reach \SI{1.065}{\watt} (Fig.~\ref{fig:highpower}(c)).
At this power level, we observe no damage to the sample or degradation in performance.
However, coupling becomes more challenging due to thermal effects, as expected.
We use a copper chipholder to improve heat dissipation, without employing active cooling.
As the pump power is increased and SH output maximized, the pump wavelength shifts from \SI{1549.3}{\nano\meter} to \SI{1553.0}{\nano\meter}.
This shift is attributed to a combination of nonlinear and thermal phase shifts, which modify the quasi-phase-matching (QPM) condition.
As our PPLT waveguides are just \SI{7}{\milli\meter}-long, we observe only minor impacts of the sample non-uniformity, and SH spectra generated by our waveguides appear close to the expected shapes (Fig.~\ref{fig:highpower}(d)).
However, they are wider than what is predicted by numerical modeling, which may indicate a certain thickness gradient, reducing the efficiency.

To further explore the potential of our PPLT waveguides, we conduct more numerical simulations, as shown in Fig.~\ref{fig:highpower}(e).
While increased efficiency improves performance, the output SH power scales more slowly with efficiency than with pump power.
At high pump powers, even low-efficiency waveguides can reach the depletion regime over relatively short lengths, while longer waveguides only help to deplete the last few percent of residual pump at the same input pump power level.
After full depletion, each additional unit of pump power is converted linearly into an equal unit of SH power.
This facilitates simpler fabrication, reduces the device footprint, and mitigates bandwidth reduction and parasitic effects such as sample non-uniformity and coincident higher-harmonic generation.
Ultimately, any improvement in conversion efficiency reduces the pump power needed to achieve high SH output levels.

\subsection*{Discussion}

By optimizing both the electrode geometry and the high-voltage pulse sequence, we have developed periodically poled LiTaO$_3$ photonic integrated waveguides capable of generating and sustaining CW SH emission at power levels exceeding \SI{1}{\watt}.
The power output is comparable to state-of-the-art LiNbO$_3$ ridge waveguides~\cite{pecheur_watt-level_2021, carpenter_cw_2020, umeki_highly_2010}, and is achieved in a shorter waveguide, preserving the bandwidth.
Our PPLT waveguides can operate in a high-power regime without any observable performance degradation or permanent damage, despite the extremely high electromagnetic field intensities resulting from strong optical mode confinement in the waveguides.

Although the achieved SHG efficiency is lower than in state-of-the-art thin-film PPLN waveguides, which can exceed \SI{2000}{\%\watt^{-1}\cm^{-2}}~\cite{chen_adapted_2024}, LiTaO$_3$ circuits can still deliver substantial CW SH output powers due to their higher optical damage threshold and robustness under elevated pump powers.
Unlike PPLN waveguides, which commonly use MgO-doped substrates to reduce the coercive field and increase the damage threshold, our PPLT devices are fabricated on undoped substrates, as MgO-doped LiTaO$_3$ wafers are not currently commercially available, to the best of our knowledge.

As this work represents one of the first demonstrations of PPLT photonic integrated circuits, we expect that further optimization of the material quality, fabrication process, and device design will lead to improved efficiency and reduced pump power requirements, while still enabling high SH output.
In particular, the availability of MgO-doped or stoichiometric LiTaO$_3$ substrates would lower the required poling voltages~\cite{kim_coercive_2002, ishizuki_periodical_2003, ishizuki_mg-doped_2008}, simplify the fabrication process, and make the technology more accessible for broader use.
Periodic poling of LiTaO$_3$ enables a range of devices, including OPOs, parametric amplifiers, quadrature-squeezed light sources, visible lasers, and self-referenced frequency combs~\cite{okawachi_chip-based_2020}.
Integrated LiTaO$_3$ frequency doublers and nonlinear converters can potentially become a part of high-power PIC-based laser systems, providing direct coherent link to visible optical bands relevant for atomic clocks and other quantum and metrology tools.

\footnotesize
\noindent \textbf{Author contributions:}
N.K. performed numerical simulations, designed PPLT waveguides, developed the periodic poling routines, and carried out SH measurements.
Z.L. optimized fabrication routines, fabricated the samples, and performed microscopic sample characterization.
N.K. prepared the manuscript with contributions from all authors. 
T.J.K supervised the work.

\noindent \textbf{Funding information:}
This work is supported by funding from the Swiss National Science Foundation (SNF) under grant number 216493 (HEROIC), by the EU Horizon Europe EIC programme under grant agreement No.~101187515 (ELLIPTIC ), and by the Swiss State Secretariat for Education, Research and Innovation (SERI).

\noindent \textbf{Acknowledgments:}
The samples were fabricated in the EPFL Center of MicroNanoTechnology (CMi) and the Institute of Physics (IPHYS) cleanroom.
Two-photon microscopy imaging was performed in the UNIL Cellular Imaging Facility (CIF).
The authors thank Chengli Wang and Zhuoya Yuan for their contribution at the early stage of the project.

\noindent \textbf{Disclosures:}
All authors declare no competing interests.

\noindent \textbf{Data availability:}
All experimental datasets and scripts used to produce the plots in this work will be uploaded to the Zenodo repository upon publication of this preprint.

\bibliography{bibliography}

\end{document}


\title{Supplementary Material to:\\Watt-level second harmonic generation in periodically poled thin-film lithium tantalate}

\author{Nikolai Kuznetsov}
\thanks{These authors contributed equally to this work.}
\affiliation{Institute of Physics, Swiss Federal Institute of Technology Lausanne (EPFL), CH-1015 Lausanne, Switzerland}

\author{Zihan Li}
\thanks{These authors contributed equally to this work.}
\affiliation{Institute of Physics, Swiss Federal Institute of Technology Lausanne (EPFL), CH-1015 Lausanne, Switzerland}

\author{Tobias J. Kippenberg}
\email[]{tobias.kippenberg@epfl.ch}
\affiliation{Institute of Physics, Swiss Federal Institute of Technology Lausanne (EPFL), CH-1015 Lausanne, Switzerland}
\affiliation{Center of Quantum Science and Engineering (EPFL), CH-1015 Lausanne, Switzerland}
\affiliation{Institute of Electrical and Micro Engineering (IEM), Swiss Federal Institute of Technology Lausanne (EPFL), CH-1015 Lausanne, Switzerland}

\maketitle

{\hypersetup{linkcolor=black}\tableofcontents}

\newpage

\renewcommand{\thefigure}{S\arabic{figure}}
\renewcommand{\theequation}{S\arabic{equation}}

\section{Periodic poling with different duty cycles of the poling electrodes}

We fabricated numerous LiTaO$_3$ samples with multiple electrodes to optimize the parameters of our poling routines (internal identifier of the high-power sample from the main text: D103\_L04\_G22\_C1\_PPLT\_WG05).
Our focus was primarily on the electrode shape and the shape of the high-voltage pulse.
For each sample, we kept the material stack and fabrication steps identical, as well as the gap between the two comb electrodes.
Further studies may reveal better poling routines based on electrodes made of another metal, or based on alternative micro-fabrication methods and tools.
We used samples from the same wafer as the sample that was optically tested, as described in the main text.
Therefore, the thickness is the same for all test samples presented here and in the following.

\begin{figure*}[htb]
	\centering
	\includegraphics[width=1\textwidth]{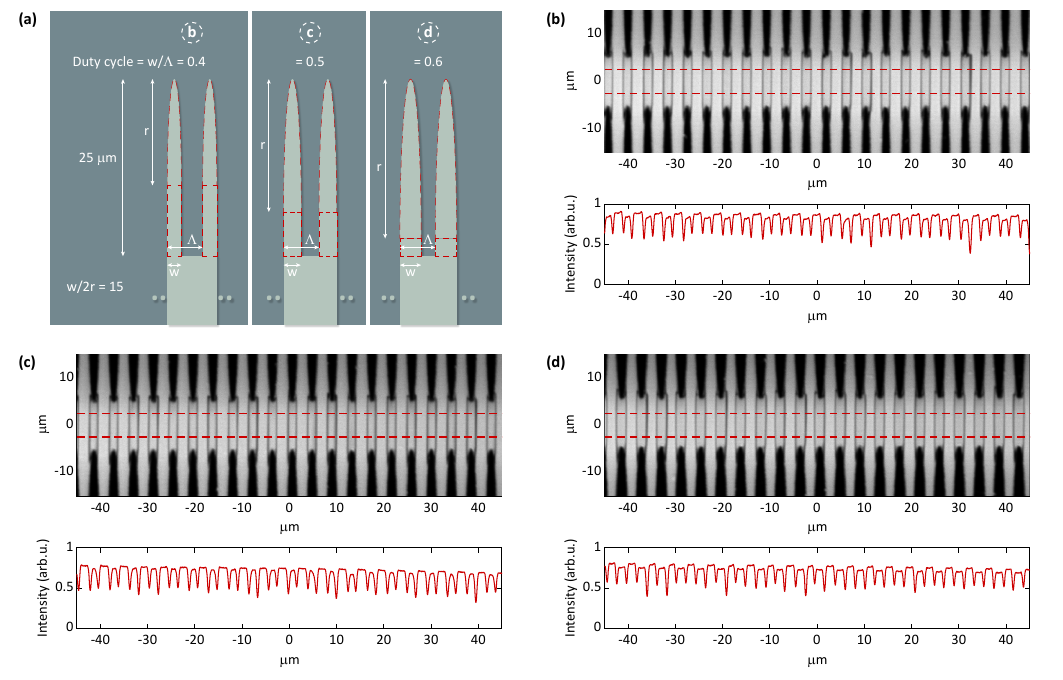}
	\caption{\textbf{Periodic poling with different duty cycles of the optimized poling electrodes.}
		\textbf{(a)} Dimensions of the electrodes used for periodic poling.
		\textbf{(b-d)} Domain inversion using the electrodes with the duty cycle of \SI{40}{\%}, \SI{50}{\%}, and \SI{60}{\%}, respectively.
		For all electrodes, the optimized train of \SI{30}{} pulses is used.}
	\label{fig:duty_cycle}
\end{figure*}

As explained in the main text, we use a two-photon microscope to verify domain inversion in LiTaO$_3$ samples.
In this and the next sections, we show images (for example, panels (b-d) in Fig.~\ref{fig:duty_cycle}) taken with this microscope to provide more insight into the poling dynamics.
Below each image, there is a panel showing the average intensity in the area between two dashed red lines.
The laser in the microscope operates at a wavelength of approximately \SI{1}{\micro\meter}, and the detector captures the emission at the respective second harmonic.
The electrodes appear as fully black periodic structures in the upper and lower parts of these images, because they do not produce any optical signal that can be captured with the microscope.
Bright areas indicate the presence of the signal generated in the thin-film LiTaO$_3$, and the thin black lines indicate the boundaries between the inverted domains.
Their black appearance is explained by the destructive interference of the photons produced in the neighboring domains.

Fig.~\ref{fig:duty_cycle} shows that the duty cycle of the poling electrodes with optimized elliptical geometry and pulse shape does not significantly affect the quality of poling.
A duty cycle of \SI{50}{\%} in the domain inversion is required for maximum efficiency of nonlinear processes in waveguides.
By selecting the duty cycle of the poling electrodes, it is possible to slightly adjust the duty cycle of the poled region; however, as shown in Fig.~\ref{fig:duty_cycle}, the difference becomes evident only with a \SI{60}{\%} duty cycle of the electrodes, while poling quality at lower duty cycles is dominated by the voltage and pulse shape.

\section{Periodic poling with different poling pulse sequences}

\begin{figure*}[htb]
	\centering
	\includegraphics[width=1\textwidth]{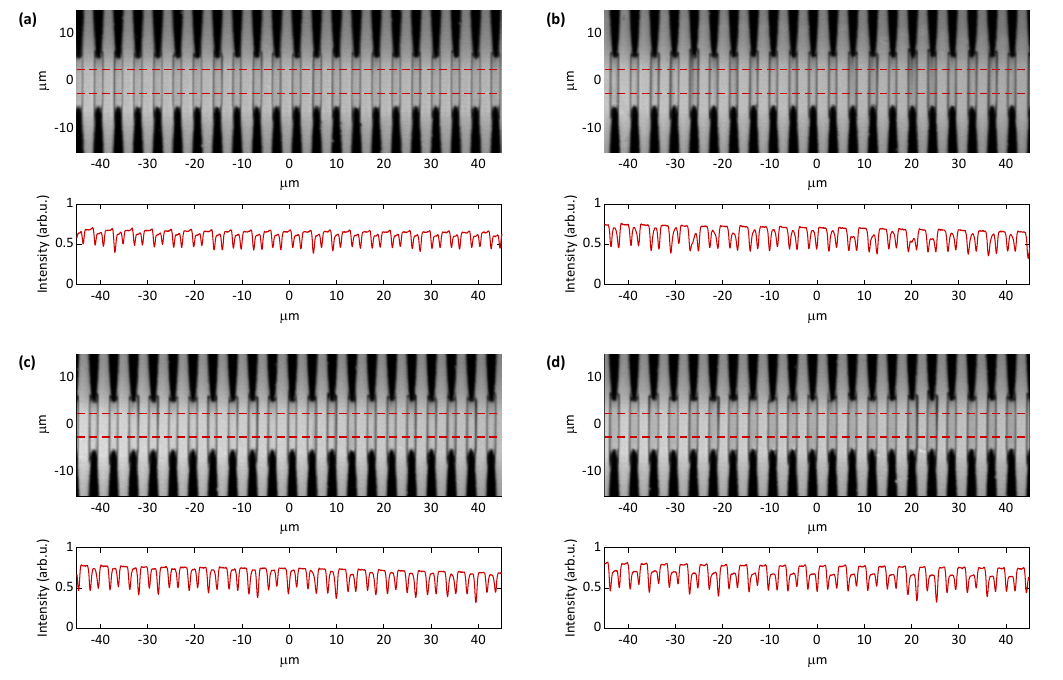}
	\caption{\textbf{Periodic poling with different poling pulse sequences.}
		\textbf{(a-d)} Poling with \SI{10}{}, \SI{20}{}, \SI{30}{}, and \SI{40}{} pulses, respectively.
		Here, the duty cycle of \SI{50}{\%} is shown for all electrodes.
		The shape of the electrodes is the optimized elliptical shape with the axis ratio of \SI{15}{}.}
	\label{fig:pulse_sequence}
\end{figure*}

The maximum voltage of the poling pulse, as well as its shape, are crucial to achieving highly uniform domain inversion.
If the maximum voltage is too low, the electric field is not strong enough to exceed the coercive field in the crystal; if the voltage is too high, the sample becomes overpoled.
After optimizing the pulse shape, we chose to use the pulse train shown in the main text.
It consists of \SI{30}{} pulses with a maximum voltage of \SI{1}{\kilo\volt}, and each pulse has a \SI{0.5}{\milli\second} rise time, \SI{0.5}{\milli\second} flat top, and \SI{0.5}{\milli\second} fall time.
The spacing between pulses is \SI{1.5}{\milli\second}.
In Fig.~\ref{fig:pulse_sequence}, we show that there is, in principle, no significant difference between poling with \SI{10}{}, \SI{20}{}, \SI{30}{}, or \SI{40}{} pulses.

\section{Quality of periodic poling with different electrode shapes}

\begin{figure*}[htb]
	\centering
	\includegraphics[width=1\textwidth]{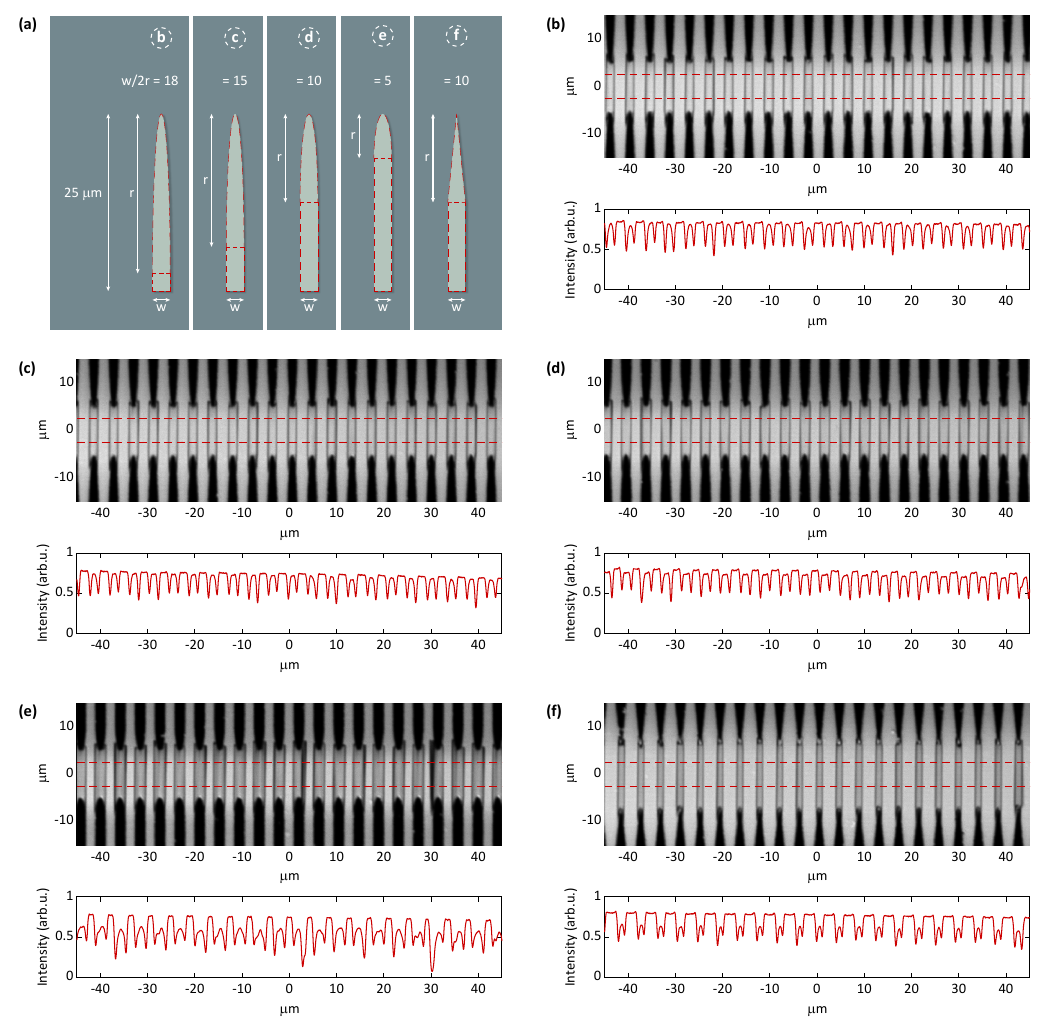}
	\caption{\textbf{Quality of periodic poling with different electrode shapes.}
		\textbf{(a)} Dimensions of different electrode tips. 
		\textbf{(b-e)} Poling with the electrode with elliptical tips with axis ratios of  
		\SI{18}{}, \SI{15}{}, \SI{10}{} and \SI{5}{}, respectively.
		\textbf{(f)} Domain inversion using the electrodes with triangular tips.
	}
	\label{fig:electrode_shape}
\end{figure*}

We found that the electrode shape significantly contributes to the final poling quality.
We tested numerous electrodes with elliptical and triangular tips, as shown in Fig.~\ref{fig:electrode_shape}(a).
Elliptical tips with axis ratios of \SI{18}{} and \SI{15}{} (Fig.~\ref{fig:electrode_shape}(b,c)) produced the best results.
Electrodes with an axis ratio of \SI{10}{} (Fig.~\ref{fig:electrode_shape}(d)) still look good, and the inverted domains are wider, bringing the duty cycle closer to the desired value.
However, the domains are slightly less uniform, leading us to prefer electrodes with an axis ratio of \SI{15}{}.
Reducing the axis ratio further consistently results in wider domain inversion and lower uniformity (Fig.~\ref{fig:electrode_shape}(e)).
The triangular tips also show good poling quality (Fig.~\ref{fig:electrode_shape}(f)); however, the duty cycle is typically below the desired value of \SI{50}{\%}.
An increase in the number of pulses or in the poling voltage results only in an insignificant lateral widening of the inverted domains.
As expected, sharper triangular tips or elongated elliptical electrodes provide better poling quality, as the cornering effect enhances the electric field at the electrode tips.

\section{Single-pulse periodic poling of thin-film lithium tantalate}

\begin{figure*}[htb]
	\centering
	\includegraphics[width=1\textwidth]{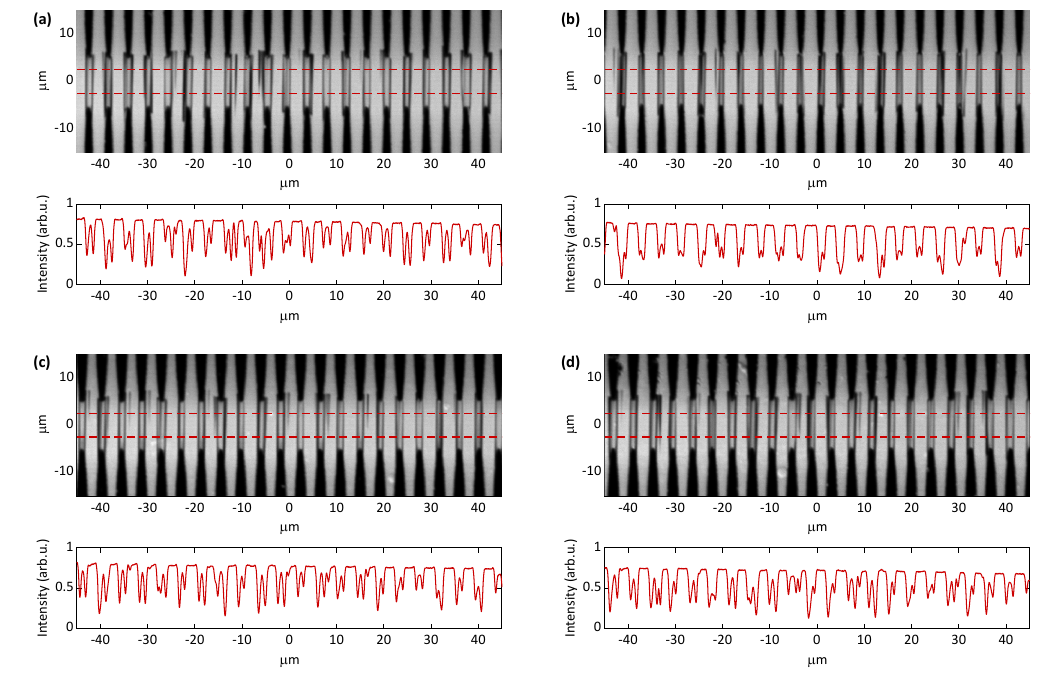}
	\caption{\textbf{Single-pulse periodic poling of thin-film lithium tantalate.}
		\textbf{(a,c)} Domain inversion quality under a single \SI{1.2}{\kilo\volt} pulse. \textbf{(b,d)} Domain inversion quality under a single \SI{1.3}{\kilo\volt} pulse.
		The images show low quality of poling.}
	\label{fig:single_pulse}
\end{figure*}

Our initial attempts to perform periodic poling of LiTaO$_3$ were based on using single pulses for domain inversion, similar to methods typically used for LiNbO$_3$.
However, we could not find an optimal strategy for achieving high-quality poling with this technique.
None of the electrode shapes or high-voltage pulse shapes we tested resulted in high-quality, reproducible domain inversion.
While the waveguides fabricated from the samples poled with single pulses show some SHG, the efficiency is low and varies between samples.
In Fig.~\ref{fig:single_pulse}, we show the results of poling with single pulses at \SI{1.2}{\kilo\volt} (left column) and \SI{1.3}{\kilo\volt} (right column).
The duty cycle is \SI{40}{\%} in Fig.~\ref{fig:single_pulse}(a,b) and \SI{60}{\%} in Fig.~\ref{fig:single_pulse}(c,d).
The inverted domains have a strongly chaotic appearance, making the single-pulse approach incompatible with our samples and fabrication methods we use.
However, the parameter space is large, and single-pulse poling might perform better with different electrodes or under different conditions defined by the material stack and fabrication steps.

\section{Cross-talk between neighboring electrodes during poling}

\begin{figure*}[htb]
	\centering
	\includegraphics[width=1\textwidth]{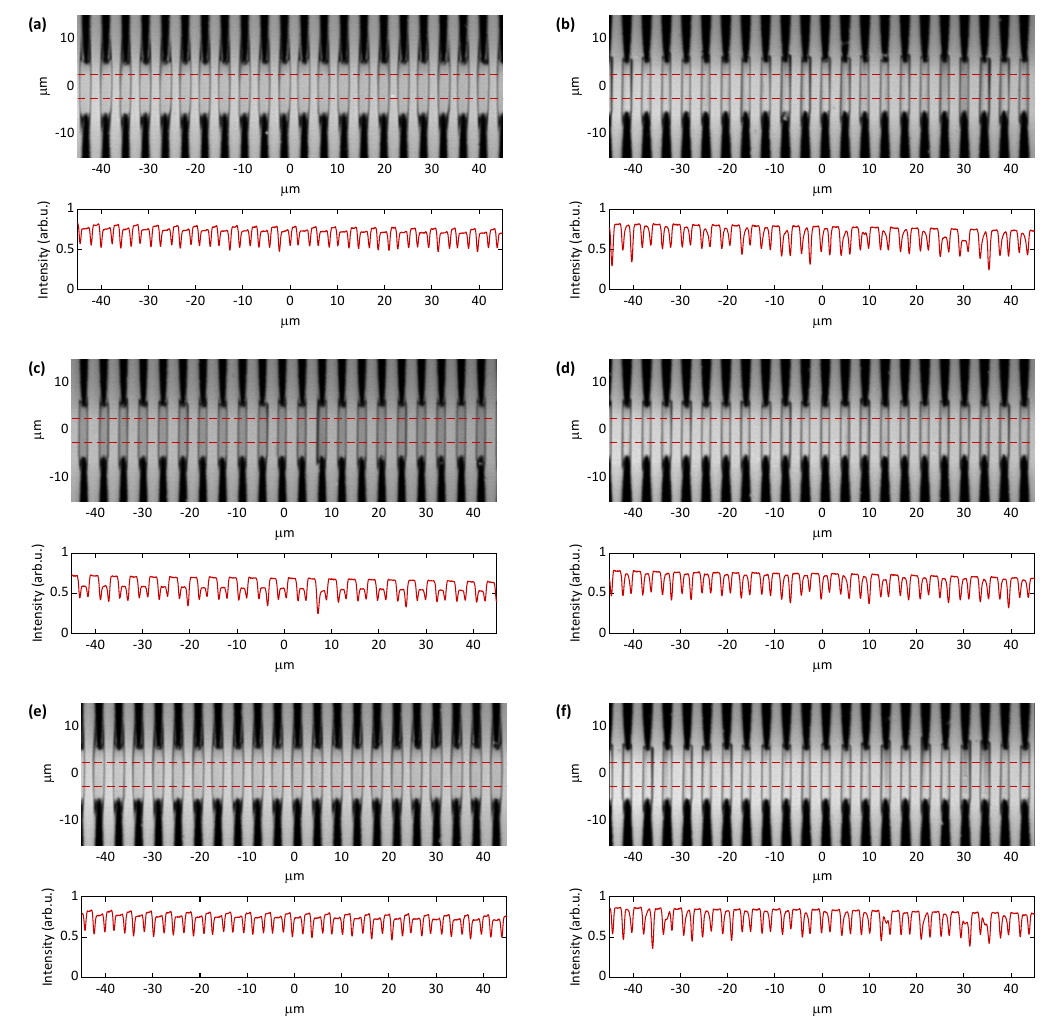}
	\caption{\textbf{Poling cross-talk between neighboring electrodes.}
		\textbf{(a,c,e)} Three regions poled under identical conditions show different domain inversion quality due to the poling cross-talk.
		\textbf{(b,d,f)} Three regions, showing better uniformity and much less affected by the poling cross-talk, as the spacing increased from \SI{40}{\micro\meter} to \SI{120}{\micro\meter}.}
	\label{fig:cross_talk}
\end{figure*}

We observed that it is important to maintain a certain distance between neighboring pairs of poling electrodes on the same photonic chip.
If the distance is too small, poling cross-talk occurs -- a situation in which the poling quality of the area between one pair of electrodes is affected by the poling process in a neighboring pair.
This is illustrated in Fig.~\ref{fig:cross_talk}.
The left column shows three poled regions spaced vertically by \SI{40}{\micro\meter}.
Fig.~\ref{fig:cross_talk}(a,e) show a clear difference compared to the poling in Fig.~\ref{fig:cross_talk}(c), although the electrodes and poling conditions are identical.

We tentatively attribute this to charge accumulation on neighboring electrodes after the electric pulse is applied to one particular pair.
Although it is challenging to collect sufficient statistical data to observe consistent patterns, we noticed that typically one region becomes overpoled if the previous one is underpoled.
However, this effect is not entirely reproducible because the time interval between two poling attempts can also influence the result.
While changes in the material stack or fabrication routines may help mitigate this issue, a simple and effective solution is to increase the distance between electrode pairs.
If the electrodes are spaced by more than \SI{120}{\micro\meter}, we typically do not observe any poling cross-talk.
The right column of Fig.~\ref{fig:cross_talk} shows three electrode pairs spaced by \SI{160}{\micro\meter}, where poling appears much more reproducible.
This observation is crucial, as poling cross-talk limits the number of devices that can be placed on the same photonic chip.

While all the testing electrodes are \SI{1}{\milli\meter} long, the electrodes used in the actual optical devices can be longer.
With our poling routines, we did not observe any difference in poling quality for electrodes of different lengths.
The presence of poling cross-talk also seems to be independent of this parameter and primarily defined by the electrode pair spacing.

\newpage
\bibliography{bibliography}